# Parallel Electric Fields And Electron Heating Observed In The Young Solar Wind

By Forrest Mozer[1,2], Oleksiy Agapitov[1], Kyung-Eu Choi[1], Richard Sydora[3]
1 Space Sciences Laboratory, University of California, Berkeley, California, USA
2 Physics Department, University of California, Berkeley, California, USA
3 Physics Department, University of Alberta, Edmonton, Canada, T6G 2E1

The largest electric fields between 18 and 30 solar radii are in narrow band waves simultaneously observed at a few Hz (somewhat above the local proton gyrofrequency) and a few hundred Hz (far below the lower hybrid frequency), with the higher frequency wave triggered at specific phases of the lower frequency wave. This wave pair, called "triggered ion acoustic waves" (TIAW), has been shown to both be physical and to occur at times of electron heating. A theory of electron heating and acceleration by the low frequency wave has been presented. While this theory and the TIAW results strongly suggest the presence of low frequency electric fields that are parallel to the local magnetic field, such fields have not been directly observed. In this paper, such parallel electric field observations are reported and TIAW are further described to conclude that they occur during about 75% of the Parker Solar Probe passes through 18 to 30 solar radii and, when present, they are the dominant wave signal, lasting for hours. In the presence of these parallel electric fields, electrons are heated while, in their absence, there is no electron heating. That there is no heating between 18 and 30 solar radii in the absence of TIAW is a most significant result because it invalidates other proposed mechanisms that predict heating in this radial range all of the time.

INTRODUCTION

"Triggered ion acoustic wave" (TIAW) is the name given to a pair of simultaneously occurring narrow band electrostatic waves whose frequencies are respectively a few Hz and a few hundred Hz, with the higher frequency wave having peak amplitudes at fixed phases of the lower frequency wave. A 12-hour continuous observation of such waves starting at a solar distance of about 20 solar radii on Parker Solar Probe orbit 7 were used to define and describe these waves that were modulated at a frequency of 1.5 Hz [Mozer et al, 2021]. Thirteen features of the TIAW later showed that they are a physical phenomenon and not an artifact arising from noise or a poor measurement [Mozer et al, 2023]. These features included whether the triggered ion-acoustic waves are electrostatic, whether both frequencies are narrowband, whether they satisfy the requirement that the electric field is parallel to the k-vector, whether the phase difference between the electric field and the density fluctuations is 90°, whether the two waves have the same phase velocity as they must if they are coupled, whether the phase velocity is that of an ion-acoustic wave, whether the electric field instrument otherwise performed as expected, etc.

From studies of orbits 6, 7, 8, and 9, it was also shown that, between 18 and 30 solar radii and in the absence of triggered ion acoustic waves, the core electron temperature, obtained from analysis of the quasi-thermal noise [Meyer-Vernet et al, 1989] decreased with increasing radial distance, R, as $R^{-4/3}$ due to the cooling produced by adiabatic expansion in the absence of heat input, while, in the presence of triggered ion acoustic waves, the core electrons were heated by as much as a factor of two [Mozer et al, 2022]. Triggered ion acoustic waves were always observed in coincidence with this core electron heating.

These data strongly suggest that a few Hz parallel electric field in the TIAW causes electron heating. A theory demonstrating such an effect was presented by Kellogg et al [2024], who showed that the strong damping of the low frequency ion acoustic waves rapidly delivers their energy to the plasma and that the resulting heating by the observed waves is not only sufficient to produce the observed electron temperature increase but it can also provide much of the outward acceleration of the solar wind.

To complete such studies, it is necessary to show that parallel electric fields at a few Hz actually exist and are observed in TIAW. This paper presents further properties of the TIAW and describes the first in-situ observation of a parallel electric field in the young solar wind. The electric field data in this work was provided by the FIELDS instruments [Bale, et al, 2016], while the electron data, excluding the core electron temperature, comes from the SWEAP instrument [Whittlesey et al [2020].

DATA

To obtain a statistically significant estimate of the frequency of TIAW occurrence between 18 and 30 solar radii, the electric field spectra near perihelion on orbits 10, 11, 12, 13, 14, and 15 are presented in Figure 1. The higher frequency TIAW are identified by the few hundred Hz narrow band electric field spectra after perihelion on orbit 10, both before and after perihelion on orbits 11 and 12, before perihelion on orbit 13, both before and after perihelion on orbit 14 and after perihelion on orbit 15. Combining this data with similar results on orbits 6 through 9 [Mozer et al, 2021] shows that TIAW occurred on 15 of 20 satellite passes between 18 and 30 solar radii. On each occurrence, they were the dominant observed electric field that lasted for hours.

To further study the relationship between the presence of TIAW and electron heating, plots of the electron temperature as a function of solar radius are presented in Figure 2 for the average of eight passes with TIAW and four passes without TIAW. The green dashed line in this figure is the $R^{-4/3}$ radial dependence of the electron temperature expected in the absence of heating and due to the cooling associated with the adiabatic expansion. From the agreement between the dashed green line and the temperature observed in the absence of TIAW, it is concluded that there was little or no electron heating in these cases. By comparison, the radial temperature dependence in the presence of the TIAW shows that electron heating occurred in this altitude range to cause the electron temperature to be significantly larger. Thus, there is a strong correlation between electron heating in the 18-30 radial range and the presence of TIAW. These are identical results to those obtained in the earlier study of orbits 6 through 9 [Mozer et al, 2022]

To close the topic of electron heating by low frequency electric fields in TIAW, it is required to show that low frequency electric fields parallel to the local magnetic field actually exist in TIAW and that they are responsible for the electron heating described by Kellogg et al, [2024]. The following discussion describes the geometry associated with such a parallel electric field observation.

The electric field on PSP is measured in the spacecraft X-Y frame, which is perpendicular to the Sun-satellite line. In spite of having only a two-component measurement, useful data for determining the parallel electric field occurs when $B_Z$, the Z-component of the magnetic field, is small such that the magnetic field is largely in the same X-Y plane as the electric field measurement. A prime coordinate system is designed to study this situation.

Its geometry is obtained by rotating every measured electric and magnetic field data point around the Z-axis through the angle whose tangent is $B_Y/B_X$, which makes $B_{Y\_prime} = 0$. The resulting geometry in the X_prime-Z coordinate system, is illustrated in Figure 3. In this frame, the angle, θ, between the magnetic field magnitude, B, and the X_prime axis has a sine that is $B_Z/B$. Thus

$$E_{X\_prime} = E_{parallel}*\cos(\theta) + E_{perpX'Z}*\sin(\theta) \qquad (1)$$

where $E_{parallel}$ is the parallel electric field, and $E_{perpX'Z}$ is the component of the perpendicular electric field in the X_prime-Z plane.

Also, the other component of the perpendicular electric field is given as

$$E_{Y\_prime} = E_{perpY'Z} \qquad (2)$$

where it is noted that $E_{perpX'z}$ is not $E_{perpY'z}$ and it is not measured.

The TIAW on inbound orbit 9 [Mozer et al, 2022, Figure 3] provides the parallel electric field example in the prime reference frame that is illustrated in Figure 4. As seen in Figure 4a, $B_Z/B$ ~0.02, such that equation (1) gives

$$E_{X\_prime} = 0.999 E_{parallel} + 0.02 E_{perpX'z} \qquad (3)$$

which means that $E_{X\_prime}$ of Figure 4b is essentially the parallel electric field at frequencies above 100 Hz and $E_{Y\_prime}$ of Figure 4c is one component of the perpendicular electric field. Because $E_{X\_prime}$ is about a factor of four larger than $E_{Y\_prime}$, it is a reliable measurement of the parallel electric field. Because it is modulated at a frequency of about 6 Hz it is clearly the higher frequency component of a TIAW event. The fact that the perpendicular electric field also shows a small component of the same modulation may be because the wave propagation was not exactly aligned with the background magnetic field or it may arise from a few degrees uncertainty of the rotation into prime coordinates. Figures 4e and 4f give a short segment of the higher frequency wave that show it to be a nearly monochromatic signal, as is typical of TIAW. Its frequency of about 250 Hz is below the lower hybrid frequency of 6000 Hz.

Figure 5 presents the 2-10 Hz band pass filtered electric field signature at the time of interest. The parallel electric field of Figure 5b describes the lower frequency TIAW wave at a frequency of about 6 Hz, which is somewhat larger than the proton gyrofrequency of 4 Hz. It is also a narrow band signal. Its parallel electric in Figure 5b has the same temporal structure as the higher frequency wave of Figure 4, which shows that the two waves are coupled with the higher frequency wave triggered at a fixed phase of the lower frequency wave. The perpendicular electric field component of Figure 5c is smaller than the parallel component of Figure 5b and its waveform has a different temporal structure. This suggests that there may be a low frequency electromagnetic signal as well as the observed electrostatic wave during this event.

CONCLUSIONS

The above discussion shows that;
1. Triggered ion acoustic waves occur between 18 and 30 solar radii during 75% of the Parker Solar Probe passes and, when present, they last for hours.
2. The triggered ion acoustic wave is the dominant wave when it is present.

3. Parallel electric fields at both a few Hz and at a few hundred Hz are directly measured in a TIAW.
4. That the parallel electric field is the major field component is further proof that they are ion acoustic waves traveling nearly parallel to the background magnetic field.
5. In the presence of these parallel electric fields, electrons are heated while, in their absence, there is little or no electron heating between 18 and 30 solar radii.

That there is little heating between 18 and 30 solar radii in the absence of TIAW is a most significant result because it invalidates other possible heating mechanisms, for example, Boldyrev et al., [2020], that produce heating in the 18-30 radial range on all passes. Thus, to our knowledge, TIAW electron heating is the only mechanism that is consistent with the observed heating and non-heating of electrons near the Sun.

ACKNOWLEDGEMENTS

This work was supported by NASA Contract Numbers NNN06AA01C and 80NSSC21K0581. The authors acknowledge the extraordinary contributions of the Parker Solar Probe spacecraft engineering team at the Applied Physics Laboratory at Johns Hopkins University. The FIELDS experiment on the Parker Solar Probe was designed and developed under NASA Contract No. NNN06AA01C. Our sincere thanks to P. Harvey, K. Goetz, and M. Pulupa for managing the spacecraft commanding, data processing, and data analysis, which has become a heavy load thanks to the complexity of the instruments and the orbit. We also acknowledge the SWEAP team for providing plasma data.

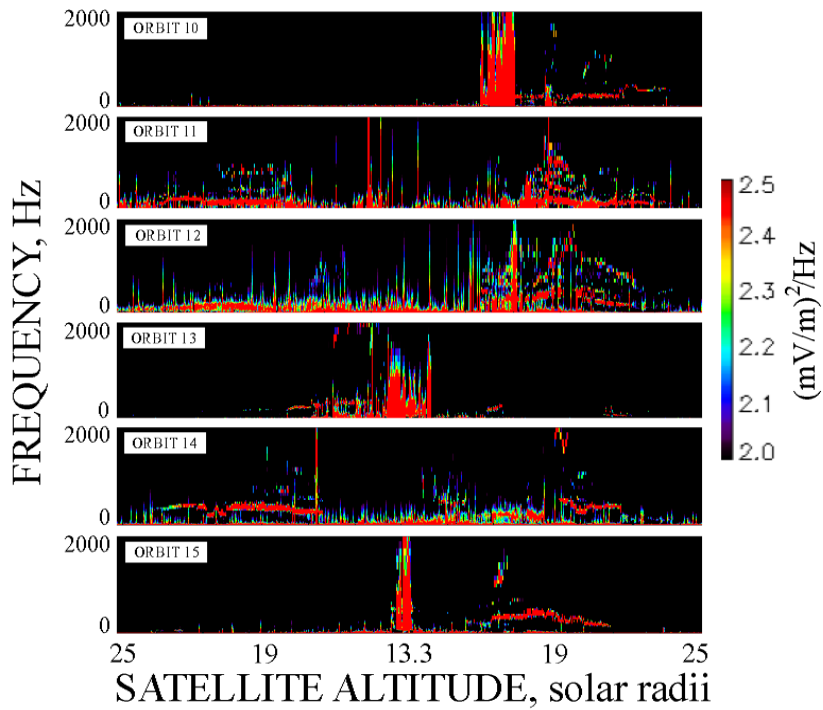

Figure 1. Electric field spectra measured near six perihelion passes of the Parker Solar Probe. The narrow band, few hundred Hz, many hours duration signals during nine of the twelve passes between 18 and 25 solar radii are the higher frequency components of triggered ion acoustic waves.

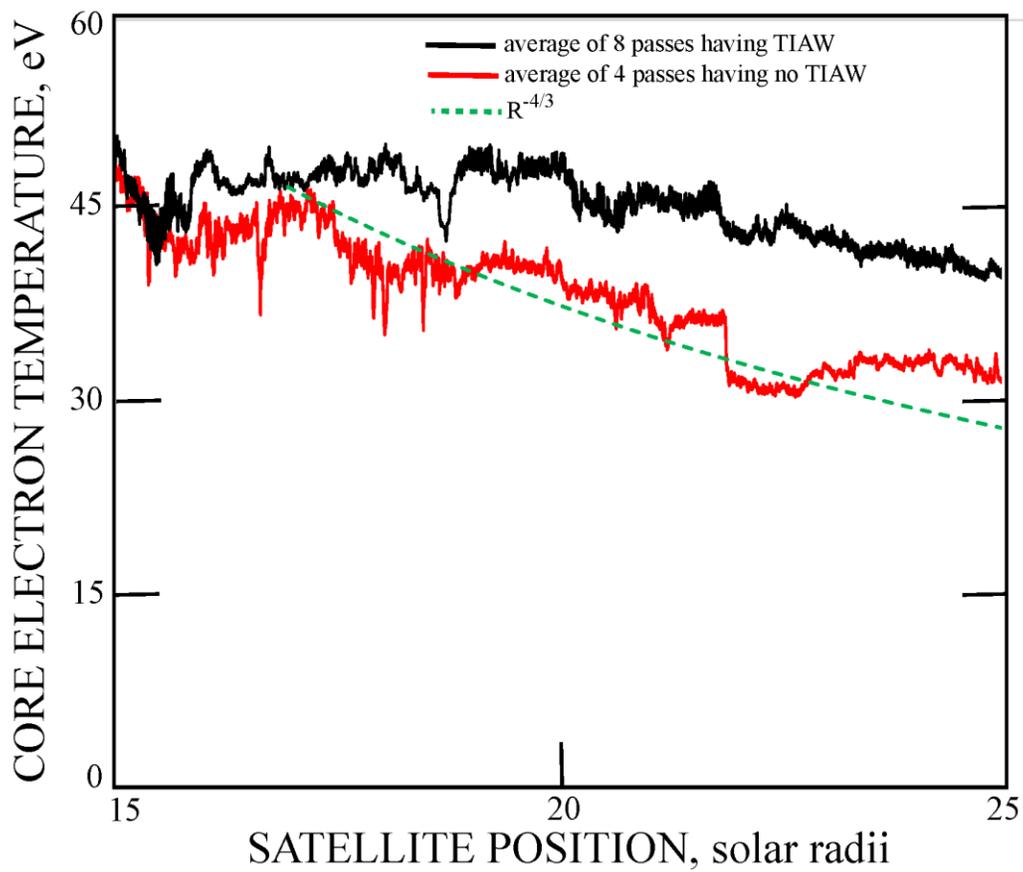

Figure 2. Core electron temperatures as a function of solar radius for eight satellite passes having triggered ion acoustic waves and four passes without such waves. The electron temperature in the absence of TIAW is less than that in the presence of TIAW and it decreases with increasing radius due to cooling associated with the adiabatic expansion. The green dashed curve is the temperature variation expected for this cooling in the absence of heating.

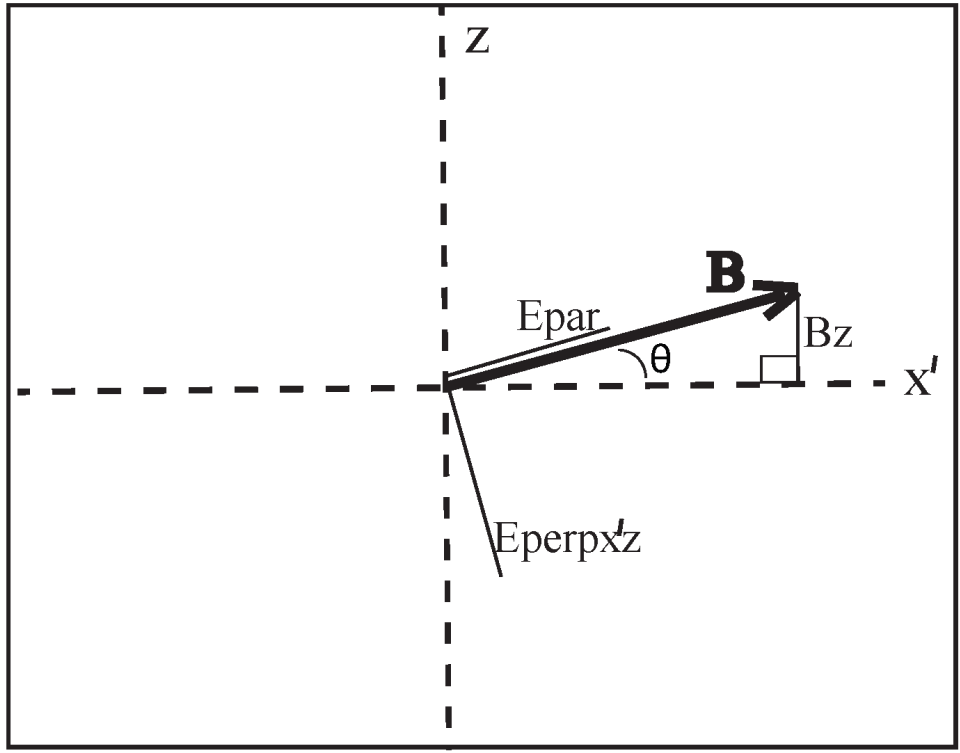

Figure 3. Geometry of the prime coordinate system in which $B_{Y\_prime}$ is zero, so the magnetic field is in the X_prime-Z plane.  In this frame $EX\_prime = E_{PARALLEL}*\cos(\theta) + E_{PERPX'Z}*\sin(\theta)$, where $\sin(\theta) = B_Z/B$.

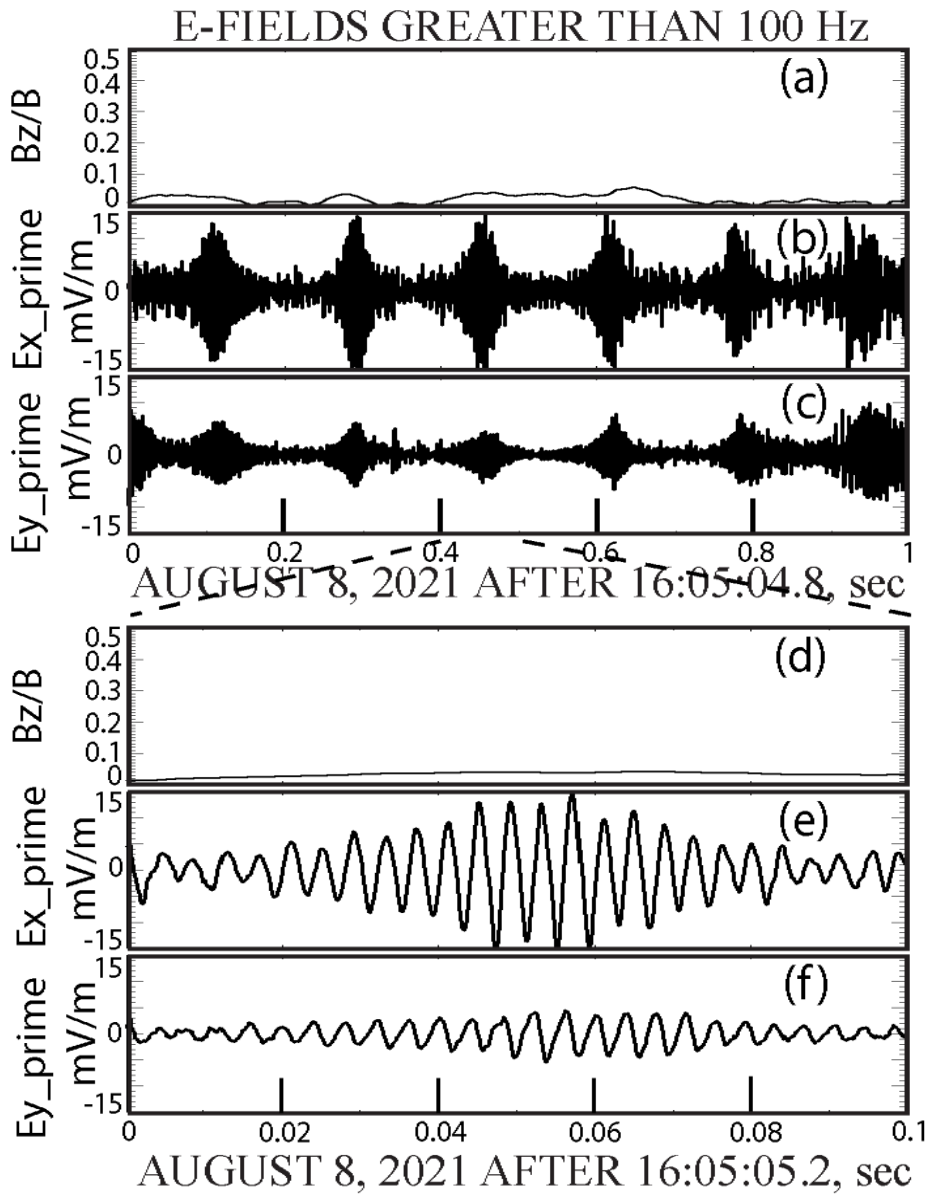

Figure 4. A one-second interval of a triggered ion acoustic wave during which B$_z$ was essentially zero in Figure 4a, such that EX_prime of Figure 4b was the >100 Hz parallel electric field, modulated at 6 Hz. An expanded section of this wave in Figure 4e shows that it was a narrow band wave at 250 Hz.

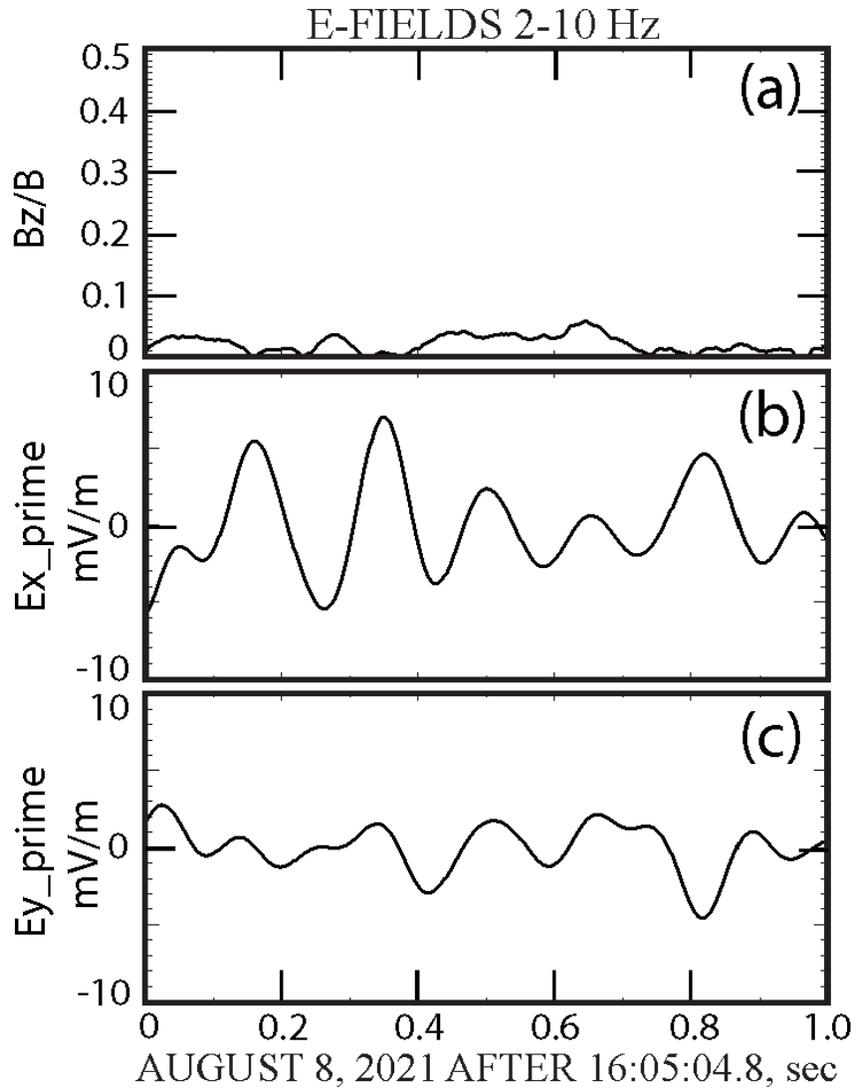

Figure 5. The low frequency triggered ion acoustic wave. Figure 5b gives the parallel component of this wave because $B_z$ of Figure 5a was small. The frequency of this wave is the same as the modulating frequency of the higher frequency wave of Figure 4, which shows that the higher frequency wave was coupled with and modulated at the frequency of the lower frequency wave.

10